\documentclass[%
reprint,
amsmath,amssymb,
aps,
pre,
]{revtex4-2}

\usepackage{graphicx}
\usepackage{mathptmx}
\usepackage{bm}
\usepackage[colorlinks=true]{hyperref}
\usepackage{CJK}
\usepackage{appendix}
\usepackage{color}

\newcommand{\ba}[1]{\bm{a}^{(#1)}}

\newcommand{\bd}[1]{\bm{d}^{(#1)}}%
\newcommand{\bbd}[1]{\bar{\bm{d}}^{(#1)}}
\newcommand{\bba}[1]{\bar{\bm{a}}^{(#1)}}

\newcommand{\bdelta}{\bm{\delta}}
\newcommand{\beeta}{\bm{v}}
\newcommand{\bg}{\bm{g}}

\newcommand{\bu}{\bm{u}}
\newcommand{\bv}{\bm{v}}
\newcommand{\bx}{\bm{x}}
\newcommand{\bxi}{\bm{\xi}}

\newcommand{\bH}[1]{{\cal H}^{(#1)}}
\newcommand{\bS}[1]{{\cal S}^{(#1)}}
\newcommand{\f}[1]{f^{(#1)}}
\newcommand{\pp}[2]{\frac{\partial #1}{\partial #2}}

\begin{document}

\title{Second-order force scheme for lattice Boltzmann method}
\author{$^{1}$Xuhui Li}
\email{lixuhui@hrbeu.edu.cn}
\author{$^{1}$Wenyang Duan}
\email{duanwenyang@hrbeu.edu.cn}
\author{$^{2}$Xiaowen Shan}
\email{shanxw@sustech.edu.cn}

\affiliation{ $^{1}$College of ShipBuilding Engineering, Harbin
	Engineering University, Harbin, Heilongjiang, 150001, China\\
	$^{2}$Department of Mechanics and Aerospace Engineering, Southern University of
	Science and Technology, Shenzhen, Guangdong 518055, China}
\date{\today}

\begin{abstract}
We present an \emph{a priori} derivation of the force scheme for lattice
Boltzmann method based on kinetic theoretical formulation.  We show that the
discrete lattice effect, previously eliminated \textit{a posteriori} in BGK
collision model, is due to first-order space-time discretization and can be
eliminated generically for a wide range of collision models with second-order
space-time discretization.  Particularly, the force scheme for the recently
developed spectral multiple-relaxation-time (SMRT) collision model is obtained
and numerically verified.
\end{abstract}

\maketitle

\section{Introduction}

\label{section:I}

Since the early development of lattice Boltzmann method (LBM), implementation of
the body force term has been generating continued interests~\cite{Bawazeer2021}.
This is especially true in recent years as more complex collision models are
adopted in place of the Bhatnagar-Gross-Krook (BGK)~\cite{Bhatnagar1954} model,
and LBM applied to multiphase flows where the interaction is modeled by a Vlasov
body force~\cite{Shan1993,Shan1994a}. Both trends call for more accurate
treatment of the body force term. Literally a dozen force schemes have been
proposed and extensive numerical evaluations
conducted~\cite{Huang2011d,Sun2012}. Nevertheless, conclusion has not been
reached on how the force term should be implemented independent of the details
of the collision model and underlying lattice.

One of the earliest force schemes is the intuitive velocity-shift method used in
the modeling of inter-molecular interactions~\cite{Shan1993}. This method shifts
the velocity in the equilibrium distribution by $\bg\Delta t$, where $\bg$ is
the acceleration and $\Delta t$ the time step. A more refined
scheme~\cite{Guo2002} eliminated the discrete lattice effect by introducing
unknown coefficients in the force term and determining them by matching the
recovered macroscopic equation with Navier-Stokes equations. Benchmarks in the
context of non-ideal gas showed that with the velocity-shift scheme, the
equilibrium densities has an un-physical dependence on the relaxation time
whereas with the latter scheme this abnormality is completely
eliminated~\cite{Yu2009}. Li \textit{et al.}~\cite{Li2012b} analyzed the exact
difference method (EDM)~\cite{Kupershtokh2004}, and found that its great
stability in Shan-Chen (SC) model ~\cite{Shan1993} attributes to the extra error
introduced to the pressure tensor. Furthermore, Li \textit{et
	al.}~\cite{Li2012b} proposed an improved force scheme to deal with the
high-density ratio in the SC model. Several force schemes for the central-moment
multiple-relaxation-time collision models with standard lattice have also been
developed~\cite{Lycett-Brown2016,DeRosis2017a,Fei2017}.

It is worth noting that in terms of the hydrodynamic moments, the leading effect
of all the models are unanimously the same, namely, when considered as an
addition to the normal collision term, the zeroth moment of the body force term
vanishes to ensure mass conservation and the first moment equals to $\bg\Delta
t$.  The subtle differences are only in the second and higher moments
representing the additional momentum and energy fluxes caused by the body force.
This effect manifests in cases such as multiphase flow modeling where the
additional stress plays a significant role~\cite{Yu2009}.

The complexity and controversy are partially due to the fact that the
development of the force schemes mostly followed the same \textit{a
 posteriori} approach of the LB development. The LB equation is constructed as a
kinetic model fully discretized in both the velocity and configuration spaces
with a discretized time. The errors caused by all discretizations are then
optimized to achieve the correct macroscopic Navier-Stokes equation. Besides
being disconnected from the Boltzmann equation which includes the effect of the
body force, the multiple expansion approach quickly becomes unmanageable for
more complex collision models.

In the kinetic theoretical formulation of LB~\cite{Shan1998,Shan2006b}, the LB
equation is obtained by first discretizing the continuum Boltzmann-BGK equation
in the velocity space, resulting in a set of partial differential equations in
the normal space and time for a set of discrete-velocity distribution function.
Without any ambiguity, the body force is given by the finite Hermite expansion
of the body force term~\cite{Martys1998,Shan2006b}.  However, to further
discretize the space and time with a finite difference scheme, multiple schemes
exist.  

In the present work we first show that the scheme of Guo \textit{et
	al}~\cite{Guo2002} can be obtained \textit{a priori} by further discretizing the
discrete-velocity Boltzmann equation with the second-order finite difference in
space and time.  The same finite difference scheme can be applied to the
recently suggested SMRT collision model~\cite{Shan2019,Li2019,Shan2021}.

The LBE is obtained by further discretizing Eq.~(\ref{eq:dv}) in space and time
using first-order forward-Euler finite difference scheme. It was shown by
multiple expansions that after absorbing the leading order error into the
dissipation term, the scheme is effectively second-order with a viscosity
proportional to $\tau+1/2$ instead of $\tau$~\cite{Chen1998a}. In presence of a
body force, the discrete lattice effect can be eliminated~\cite{Guo2002}
\textit{a posteriori}. We show in this section that the same correction can be
obtained \textit{a priori} for the BGK model by integrating Eq.~(\ref{eq:dv})
using second-order quadrature rules~\cite{He1998}. In the next section, the same
technique is used to obtain the force scheme for the Hermite-space MRT model.

With the BGK collision model, the description of the collision as a
\emph{uniform} relaxation process of the distribution function towards its
equilibrium is in many cases simplistic. In a previous series of
papers~\cite{Li2020b,Shan2021,Shi2021}, the SMRT collision model was developed
where the \emph{irreducible components} of the Hermite coefficients are relaxed
separately in the reference frame moving with the fluid. These components are
the minimum tensor components that can be separately relaxed without violating
rotation symmetry.

The rest of the paper is organized as the following.  The theoretical derivation
is presented in Section~\ref{section:II} with the discrete-velocity force
scheme presented in Section~\ref{sec:dv}, the derivation of the force scheme
for BGK model in Section~\ref{sec:bgkf}, and force scheme for the SMRT
collision model in Section~\ref{sec:mrtf}.  Some numerical verification is given
in Section~\ref{section:VI} and the conclusion and discussions are given in
Sec.~\ref{section:VII}.

\section{Boltzmann-BGK equation}
\label{section:II}

\subsection{Background}

\label{sec:dv}
In kinetic theory~\cite{Chapman1970}, the evolution of the single-particle
distribution, $f(\bx, \bxi,t)$, under an external or
self-generated body force with acceleration $\bg$ is described by the Boltzmann
equation:
\begin{equation}
	\label{eq:Boltzmann-BGK}
	\pp ft + \bxi\cdot\nabla f + \bg\cdot\nabla_{\bxi} f = \Omega(f),
\end{equation}
where, $\bx$ and $\bxi$ are coordinates in physical and velocity spaces
respectively, $t$ the time, $\Omega(f)$ the collision term describing the effect
of inter-particle collision. Due to its extreme complexity, the collision term
is often simplified by models, of which the most widely used is the BGK model:
\begin{equation}
	\label{BGK}
	\Omega_{BGK}(f) = -\frac 1\tau\left[f - \f{eq}\right],
\end{equation}
where $\tau$ is a relaxation time and $\f{eq}$ the Maxwell-Boltzmann
distribution. Choosing the characteristic speed $\sqrt{k_BT_0/m_0}$ with $k_B$
the Boltzmann constant and $T_0$ and $m_0$ the reference temperature and
molecule mass, $\f{eq}$ has the dimensionless form
\begin{equation}
	\f{eq}(\bx, \bxi, t) = \frac{\rho}{(2\pi\theta)^{D/2}}
	\exp\left[-\frac{(\bxi-\bu)^2}{2\theta}\right],
	\label{Maxwellian}
\end{equation}
where $\rho$ is the density, $\bu$ the fluid velocity and $\theta\equiv T/T_0$
the temperature, all dimensionless.

The lattice Boltzmann equation was formulated as a special velocity-space
discretization of the Boltzmann equation based on two
observations~\cite{Shan2006b}.  First, in Chapman-Enskog
calculation~\cite{Huang1987}, the macroscopic hydrodynamics only depends on the
leading moments of the distribution function rather than its entirety. The
distribution function can therefore be approximated by its low-order Hermite
expansion without altering the hydrodynamics~\cite{Grad1949,Grad1949a}. This
truncation is equivalent to projecting Eq.~(\ref{eq:Boltzmann-BGK}) into a
low-order Hilbert space spanned by the Hermite polynomials. Denoting the $N$-th
order Hermite series by:
\begin{equation}
	\label{eq:truncation}
	f_N(\bx,\bxi,t) = \omega(\bxi)\sum_{n=0}^N\frac{1}{n!}\ba{n}:\bH{n}(\bxi),
\end{equation}
where $\bH{n}(\bxi)$ is the $n$-th Hermite polynomial, and
\begin{equation}
	\omega(\bxi)\equiv (2\pi)^{-D/2}\exp\left(-\xi^2/2\right),
\end{equation}
is the weight function with respect to which the Hermite polynomials are
orthogonal. The expansion coefficients:
\begin{equation}
	\label{eq:an}
	\ba{n}(\bx,t) = \int f_N(\bx,\bxi,t)\bH{n}(\bxi)d\bxi,
\end{equation}
are \textit{velocity moments} of the distribution function with
the leading few being the familiar hydrodynamic variables, $\rho$, $\bu$ and
$\theta$.

Second, any finite Hermite series is completely determined by its values on a
finite set of $\bxi_i$. Let $\{(w_i, \bxi_i), i = 1, \cdots, d\}$ be the weights
and abscissas of a $Q$-th degree Hermite quadrature such that for any $Q$-th
degree polynomial, $p(\bxi)$, we have:
\begin{equation}
	\int\omega(\bxi)p(\bxi)d\bxi = \sum_{i=0}^dw_ip(\bxi_i).
\end{equation}
The $M$-th moment of $f_N$ is then
\begin{eqnarray}
	\int f_N(\bxi)\bxi^Md\bxi &=& \int\omega(\bxi)
\left[\frac{f_N(\bxi)\bxi^M}{\omega(\bxi)}\right]d\bxi\nonumber\\
	&=& \sum_{i=0}^d\frac{w_if_N(\bxi_i)\bxi_i^M}{\omega(\bxi_i)},
\end{eqnarray}
provided that $Q\geq M+N$, as the integrand in the brackets is a polynomial of
degree $M+N$.  Hence, all expansion coefficients are completely determined by
$f_N(\bxi_i)$ as long as $\{\bxi_i\}$ forms the abscissas of a quadrature rule
of a degree $Q\geq 2N$.  If we further define the convenience variable:
\begin{equation}
	\label{eq:fi}
	f_i\equiv \frac{w_if_N(\bx, \bxi_i, t)}{\omega(\bxi_i)},
\end{equation}
the integral velocity moments has the discrete form:
\begin{equation}
	\int f_N(\bxi)\bxi^Md\bxi = \sum_{i=0}^df_i\bxi_i^M,
\end{equation}
provided that the quadrature conditions are met.  Noting by Eq.~(\ref{eq:an})
that the expansion coefficients are also velocity moments, $\ba{n}$ and $f_i$
can be transformed through the following general discrete Fourier transform:
\begin{subequations}
\begin{eqnarray}
	\label{eq:dft}
	\ba{n} &=& \sum_{i=0}^df_i\bH{n}(\bxi_i),\\
	f_i &=& w_i\sum_{n=0}^N\frac{1}{n!}\ba{n}:\bH{n}(\bxi_i).
\end{eqnarray}
\end{subequations}

The dynamic equations of $f_i$ are taken as the direct evaluation of the
\textit{projected} Eq.~(\ref{eq:Boltzmann-BGK}) at $\bxi_i$. This amounts to
expanding all terms in terms of Hermite polynomials and truncating to a finite
order. The expansion of $\nabla_{\bxi}f_N$ can be obtained by taking the
velocity-space derivative of Eq.~(\ref{eq:truncation}) and using the
following Rodrigues formula:
\begin{equation}
	\bH{n}(\bxi) = \frac{(-1)^n}{\omega(\bxi)}\nabla^n_{\bxi}\omega(\bxi),
\end{equation}
to write:
\begin{eqnarray}
	\label{eq:force_expansion}
	\lefteqn{\nabla_{\bxi} f_N = \sum_{n=0}^N\frac{1}{n!}
	\ba{n}:\nabla_{\bxi}\left[\omega(\bxi)\bH{n}(\bxi)\right]}\nonumber \\
	&=& -\omega(\bxi)\sum_{n=0}^N\frac{1}{n!}\ba{n}:\bH{n+1}(\bxi)\nonumber\\
	&=& -\omega(\bxi)\sum_{n=1}^{N+1}\frac{1}{n!}\left[n\ba{n-1}\right]:\bH{n}(\bxi).
\end{eqnarray}
The body-force term, denoted by $F(\bxi)\equiv -\bg\cdot\nabla_{\bxi}f_N$, has
thus the following expansion:
\begin{equation}
	\label{eq:fxi}
	F(\bxi)= \omega(\bxi)\sum_{n=1}^N\frac{1}{n!}\left[n\bg\ba{n-1}\right]:\bH{n}(\bxi),
\end{equation}
Denoting the Hermite coefficients of $F(\bxi)$ by $\ba{n}_F$, we
have
\begin{equation}
	\label{eq:af}
	\ba{0}_F = 0, \quad\mbox{and}\quad
	\ba{n}_F = n\bg\ba{n-1}, \quad n\geq 1.
\end{equation}
where $\bg\ba{n-1}$ is to be understood as the \textit{symmetric product}
between $\bg$ and $\ba{n-1}$, \textit{e.g.}, in component form, $\bg \ba{2} =
(g_{\alpha}a_{\beta\gamma} + g_{\beta}a_{\gamma\alpha} +
g_{\gamma}a_{\alpha\beta})/3$.

The explicit expressions for the first several orders of $\ba{n}_0$ can be found
in the literature ~\cite{Shan2006b}. Thus we have the expanded force term up to
the fourth-order:
\begin{eqnarray}
\label{eq:Fr}
F \thickapprox \omega(\bxi) \rho \lbrace \underbrace{\bg \cdot \bxi}_{1st} + \underbrace{(\bg \cdot \bxi)(\bxi \cdot \bu) - \bg \cdot \bu}_{2nd} \nonumber \\
+  \underbrace{ \frac{1}{6\rho}3\bg\left[\rho(\bu^2+(\theta-1)\bdelta)+\ba{2}_1  \right]:\bH{3}(\bxi)}_{3rd}\nonumber\\ 
+ \underbrace{\frac{1}{24\rho}4\bg(\ba{3}_0+\ba{3}_1):\bH{4}(\bxi)}_{4th} \rbrace \qquad
\end{eqnarray}

If the expansion of the force term is truncated to second order, the familiar
force term can be obtained:
\begin{equation}
	F(\bxi)=\rho\omega(\bxi)\left[\bg\cdot(\bxi-\bu)
	 + (\bg\cdot\bxi)(\bu\cdot\bxi)\right].
\end{equation}

Similar to Eq.~(\ref{eq:fi}), defining:
\begin{equation}
	F_i\equiv \frac{w_iF(\bxi_i)}{\omega(\bxi_i)},
\end{equation}
the discrete-velocity Boltzmann equation with body force is:
\begin{equation}
	\label{eq:dv}
	\pp {f_i}t + \bxi_i\cdot\nabla f_i = \Omega_i + F_i, \quad i = 1, \cdots, d.
\end{equation}
The above equation was previously given by Martys \textit{et
	al}~\cite{Martys1998} and its relations to the previous models were also
analyzed. Naturally the body force term is independent of the collision term,
and other than the truncation order, there is no ambiguity.

\subsection{force in BGK collision model}

\label{sec:bgkf}

The LBE is obtained by further discretizing Eq.~(\ref{eq:dv}) in space and time
using first-order forward-Euler finite difference scheme. It was shown by
multiple expansions that after absorbing the leading order error into the
dissipation term, the scheme is effectively second-order with a viscosity
proportional to $\tau+1/2$ instead of $\tau$~\cite{Chen1998a}. In presence of a
body force, the discrete lattice effect can be eliminated~\cite{Guo2002}
\textit{a posteriori}. We show in this section that the same correction can be
obtained \textit{a priori} for the BGK model by integrating Eq.~(\ref{eq:dv})
using second-order quadrature rules~\cite{He1998,Malaspinas2009,Li2022}. In the
next section, the same technique is used to obtain the force scheme for the
Hermite-space MRT model.

The discrete-velocity BGK model can be written as:
\begin{equation}
\label{eq:collision_BGK}
	\Omega_i = -\frac{1}{\tau}\left[f_i - f_i^{(eq)}\right],
\end{equation}
where:
\begin{equation}
	\f{eq}_i = \rho w_i\left\{1 + \bu\cdot\bxi + 
	\frac 12\left[(\bu\cdot\bxi)^2-u^2\right]\cdots\right\}
\end{equation}
is the truncated Hermite expansion of the Maxwellian evaluated at
$\bxi_i$~\cite{Shan2006b}.
Integrating Eq.~(\ref{eq:dv}) along the characteristic line and deal with the integral on the right-hand side by the trapezoidal rule,
we have 
\begin{eqnarray}
\label{eq:implicityeq}
f_i(\bx + \bxi,t+1) - && f_i(\bx,t) = \frac{1}{2}[\Omega_i(\bx+\bxi,t+1)+\Omega_i(\bx,t)]  \nonumber \\
                                   &&+ \frac{1}{2}[F_i(\bx+\bxi,t+1)+F_i(\bx,t)]
\end{eqnarray}
in which the time step $\Delta t = 1$ is applied for brevity.

Define a new distribution function
\begin{equation}
\label{eq:newf}
\bar{f}_i = f_i - \frac{1}{2}\Omega_i - \frac{1}{2} F_i
\end{equation}

Apply the new defined distribution function, the implicit evolution equation Eq.(~\ref{eq:implicityeq}) can be reconstructed as an explicit evolution equation
\begin{equation}
\label{eq:evolution}
\bar{f}_i(\bx + \bxi,t+1) = \bar{f}_i(\bx,t) + \Omega_i(\bx,t) + F_i(\bx,t) 
\end{equation}

Substitute the BGK collision term, Eq.~(\ref{eq:collision_BGK}), into the above
equation and replace $f$ by $\bar{f},f^{(eq)}$and $F_i$ with
Eq.~(\ref{eq:newf}), the above evolution equation can be written as the
following complete explicit form
\begin{eqnarray}
\label{eq:explict_eq}
\bar{f}_i(\bx + \bxi_i, t+1) = &&\bar{f}_i(\bx, t)  -\frac{1}{\hat{\tau}}\left[ \bar{f}_i(\bx,t) - f_i^{(eq)}(\bx,t) \right]  \nonumber \\
                               &&+ (1 - \frac{1}{2\hat{\tau}})F_i(\bx,t)
\end{eqnarray}
with $\hat{\tau}=\tau + \frac{1}{2}$. It should be noted that the equilibrium
and force term in the above equation are the original forms and $\bar{f}_i$ is
the actual distribution function in the numerical implementation.

The zero-order, first-order and second-order moments of the new defined
distribution function can be evaluated according to the original
one~\cite{Li2022}. The zero-order moment is as followings
\begin{equation}
\sum \bar{f}_i = \sum (f_i + \frac{1}{2\tau}f_i^{(1)} -\frac{1}{2}F_i) = \sum f_i
\end{equation}
in which the zero-order moments of the nonequilibrium and force term are null.
Thus we have
\begin{equation}
\bar{\rho} = \rho 
\end{equation}

The first-order moment of the new distribution function is as followings
\begin{eqnarray}
\sum \bar{f}_i \xi_{i,\alpha} &=& \sum {(f_i + \frac{1}{2\tau}f_i^{(1)} -\frac{1}{2}F_i) \xi_{i,\alpha}} \nonumber \\
                             &=& \sum {(f_i  -\frac{1}{2}F_i) \xi_{i,\alpha}}
\end{eqnarray}
in which the first-order moment of the nonequilibrium is null.
Thus we have
\begin{equation}
\label{eq:half_F}
\rho \bar{\bu} = \rho \bu - \frac{\bm{F}}{2}
\end{equation}

Then the physical velocity can be written as
\begin{equation}
\bu = \bar{\bu} + \frac{\bm{F}}{2\rho}
\end{equation}

The second-order central-moment of the new defined distribution function is as followings
\begin{eqnarray}
\sum \bar{f}_i c_{i,\alpha}c_{i,\beta} &=& \sum {(f_i + \frac{1}{2\tau}f_i^{(1)} -\frac{1}{2}F_i) c_{i,\alpha}}c_{i,\beta} \nonumber \\
                                       &=& \sum {(f_i + \frac{1}{2\tau}f_i^{(1)} ) c_{i,\alpha}}c_{i,\beta}
\end{eqnarray}

The trace of the second-order tensors in the above equation can be obtained by
contraction of the subscript index
\begin{equation}
D\rho\bar{\theta} = D\rho \theta
\end{equation}  
in which the trace of the shear stress is null. This indicates that the physical
temperature is equal to the contraction of the second-order central-moment of
the new distribution function
\begin{equation}
\theta=\bar{\theta}  
\end{equation}  

It is worth noting that the macroscopic variables used in the equilibrium and
force term in Eq.~(\ref{eq:explict_eq}) are the physical variables
$\rho,\bu,\theta$, instead of $\bar{\rho},\bar{\bu},\bar{\theta}$. In general,
the force scheme in the evolution equation of the distribution function is
derived via a rigorous and priori approach here. If thermohydrodynamic level is
not considered and the force term is only expanded to second-order, it is
reduced to the force scheme derived by Guo \textit{et al.}~\cite{Guo2002}.
\subsection{force in SMRT collision model}

\label{sec:mrtf}

The force scheme in the raw-moment Hermite MRT collision model is organized in
the appendix for the interested readers. We now apply the same technique to
derive the space-time discretization for the central-moment based SMRT collision
model~\cite{Shan2019,Li2019,Shan2021}. Briefly, the expansion of
Eq.~(\ref{eq:truncation}) is made in the reference frame moving with fluid.
Namely the Hermite polynomials are with respect to $\bv\equiv (\bxi - \bu) /
\sqrt{\theta}$ as:
\begin{equation}
	f_N(\bx, \bv, t) = \omega(\bv)\sum_{n=0}^N\frac 1{n!}\bd{n}:\bH{n}(\bv),
\end{equation}
with $\bd{n}(\bx, t)$ being the expansion coefficients given by:
\begin{equation}
	\bd{n}(\bx,t) = \int f_N(\bx,\bv,t)\bH{n}(\bv)d\bv.
\end{equation}
Note that this is precisely the Hermite polynomials used by
Grad~\cite{Grad1949a}.  The Maxwell-Boltzmann equilibrium distribution is
related to the weight function by:
\begin{equation}
	\f{0}(\bv) = \rho\theta^{-D/2}\omega(\bv).
\end{equation}
Its Hermite expansion coefficient, denoted by $\bd{n}_0$, are hence:
\begin{equation}
	\bd{n}_0 = \left\{\begin{array}{ll}
		\rho\theta^{-D/2}, & n=0   \\
		0,                 & n > 0
	\end{array}\right..
\end{equation}
Owing to the conservation of mass and momentum, we have $\bd{0} = \bd{0}_0$ and
$\bd{1} = \bd{1}_0 = 0$.  Similar to Eq.~(\ref{eq:fxi}), the expansion of
the body-force term in the rescaled central moment (RCM) space is as followings:
\begin{eqnarray}
F = \frac{\omega(\beeta)}{\sqrt{\theta}}  \sum_{n=1}^{N} \frac{1}{n!} [n \bg (\bd{n-1}_0 + \bd{n-1}_1)]:\bH{n}(\beeta). 
\end{eqnarray}
in which the relation $\nabla_{\bxi} =  \frac{1}{\sqrt{\theta}}\nabla_{\beeta}$ is applied. The detailed expansion form for the first several orders is 
as followings
\begin{eqnarray}
\label{eq:Frc}
F \thickapprox \frac{1}{\sqrt{\theta}}\omega(\beeta) \rho \lbrace \underbrace{\theta^{-D/2}\bg \cdot \beeta}_{1st} + \underbrace{0}_{2nd} \nonumber 
\qquad\qquad  \\
+  \underbrace{ \frac{1}{6\rho}\theta^{-(D+2)/2}3\bg\ba{2}_1  :\bH{3}(\beeta)}_{3rd} \qquad\nonumber\\ 
+ \underbrace{\frac{1}{24\rho}\theta^{-(D+3)/2}4\bg(\ba{3}_1-3\bu\ba{2}_1):\bH{4}(\beeta)}_{4th} \rbrace
\end{eqnarray}

If the first two terms are retained, then we have
\begin{eqnarray}
F && \thickapprox \frac{1}{\sqrt{\theta}}\omega(\beeta) \rho\theta^{-D/2}\bg \cdot \beeta  \nonumber \\
&& = \frac{1}{\sqrt{\theta}}\bg \cdot \beeta f^{eq}  = -\bg \cdot \nabla_{\bxi} f^{eq},
\end{eqnarray}
which is identical to the force scheme developed by He \textit{et
	al.}~\citep{He1998}.

For the purpose of convenience, we define the Hermite coefficients of the force
term as
\begin{equation}
	\bd{n}_F = \frac{1}{\sqrt{\theta}} n\bg\bd{n-1}, \quad n\geq 1.
\end{equation}
In particular, we have $\bd{1}_F = \rho\bg\theta^{-(D+1)/2}$ and $\bd{2}_F = 0$.
Furthermore, if the contributions from the nonequilibrium are neglected, we have
$\bd{n}_F = 0,n \geq 2$ and only $\bd{1}_F$ is not null.

To allow maximum flexibility while preserving rotational
symmetry~\cite{Li2020b,Shan2021,Shi2021}, each $\bH{n}(\bv)$ is further
decomposed into its traceless components, $\bS{n,k}(\bv)$. Let the distribution
function has the following expansions:
\begin{eqnarray}
	f(\bv) &=& \omega(\bv)\sum_{n=0}^N\frac{1}{n!}\sum_k\bd{n,k}:\bS{n,k}(\bv),
\end{eqnarray}
and $\bd{n,k}_\Omega$ the coefficients of the similar expansion of the collision
operator which is defined as the independent relaxation of each traceless
components, \textit{i.e.}:
\begin{eqnarray}
	\label{eq:domg}
	\bd{n,k}_\Omega &=& -\frac{1}{\tau_{nk}}(\bd{n,k}-\bd{n,k}_0), \nonumber \\
	                &=& -\frac{1}{\tau_{nk}} \bd{n,k}  \quad n = 1, \cdots, N,
\end{eqnarray}
where $\tau_{nk}$ are relaxation times. 

In the face of this complicated collision operator, the technique of the
previous section can still be applied. Eq.~(\ref{eq:dft}) indicates the
transform from the phase space to the raw moment (RM) space. Similarly, we
denote the transform from the phase space to the RCM space, \textit{i.e.}, $f_i$
to $\bd{n,k}$ as:
\begin{equation}
	M_{nk}(f_i) = \bd{n,k}.
\end{equation}
By Eq.~(\ref{eq:domg}), we have:
\begin{equation}
	M_{nk}(\Omega_i) = -\frac{1}{\tau_{nk}}\bd{n,k}.
\end{equation}

Starting from Eq.~(\ref{eq:dv}), we can get the same evolution equation as
Eq.(~\ref{eq:evolution}) for the SMRT model but with different collision term.
The transform of the right-hand-side of Eq.(~\ref{eq:evolution}) to the RCM
space is
\begin{equation}
\label{eq:RHS}
\bar{\bm{d}}_p^{(n,k)} = \bar{\bm{d}}^{(n,k)} - \frac{1}{\tau_{nk}}(\bd{n,k}-\bd{n,k}_0) + \bd{n,k}_F,n \geq 1
\end{equation} 
in which the subscript \textit{p} denotes the post-collision state and
$\bar{\bm{d}}^{(n,k)}$ is the mapping of Eq.~(\ref{eq:newf}) in RCM space
\begin{equation}
\bar{\bm{d}}^{(n,k)} = \bd{n,k} + \frac{1}{2\tau_{nk}}(\bd{n,k}-\bd{n,k}_0) -\frac{1}{2} \bd{n,k}_F
\end{equation}
Specifically, $\bar{\bm{d}}^{(1,k)} = - \frac{1}{2} \bd{1,k}_F $.

Thus Eq.(~\ref{eq:RHS}) can be written as the following explicit form
\begin{equation}
\bar{\bm{d}}_p^{(n,k)} = (1 - \frac{1}{\hat{\tau}_{nk}})\bar{\bm{d}}_1^{(n,k)} 
	+ (1- \frac{1}{2\hat{\tau}_{nk}})\bd{n,k}_F,\quad n\geq 1
\end{equation}
in which $\hat{\tau}_{nk} = \tau_{nk} + \frac{1}{2}$ and $\bar{\bm{d}}_1^{(n,k)}
= \bar{\bm{d}}^{(n,k)} - \bm{d}_0^{(n,k)} $. For $n=1$, we have
$\bar{\bm{d}}_p^{(1,k)} =  \frac{1}{2} \bd{1,k}_F $.

To avoid the interpolation operation in the stream process, the post-collision
RCMs need to be transferred to the RM space and then reconstruct the
post-collision distribution function. Similar to the relations between the RCMs
and RMs of nonequilibrium as \cite{Li2019,Li2022}, the transformation for the first
several orders are as followings
\begin{subequations}
\label{eq:bbd1}
\begin{eqnarray}
\bba{0}_p &=& 0, \\
\bba{1,k}_p &=& \theta^{\frac{D+1}{2}} \bbd{1,k}_p = \frac{\bm{F}}{2},   \\
\bba{2,k}_p &=& \theta^{\frac{D+2}{2}}\bbd{2,k}_p + 2\bu \bba{1,k}_p  \nonumber \\
            &=& (1 - \omega_2)\bba{2,k}_1 + (1- \frac{\omega_2}{2})\bba{2,k}_F, \\
\bba{3,k}_p &=& \theta^{\frac{D+3}{2}}\bbd{3,k}_p + 3\bu\bba{2,k}_p - 3\bu_1 \bba{1,k}_p  \\
            &=& (1 - \omega_3)(\bba{3,k}_1 - 3\bu \bba{2,k}_1 + 3\bu_1 \bba{1,k}_1)  \nonumber \\
            & &  + 3\bu\bba{2,k}_p - 3\bu_1 \bba{1,k}_p, \nonumber \\
            &=& (1 - \omega_3)\bba{3,k}_1 + (\omega_3 - \omega_2)3\bu \bba{2,k}_1  \nonumber \\
            & & + (1-\frac{\omega_3}{2})\bba{3,k}_F + (\frac{\omega_3}{2} - \frac{\omega_2}{2})3\bu \bba{2,k}_F, \nonumber \\
\bba{4,k}_p &=& 4\bu\bba{3,k}_p -6\bu_1\bba{2,k}_p  + 4\left( \bu\bu_2 \right)  \bba{1,k}_p.
\end{eqnarray}
\end{subequations}
in which $\bba{n,k}_p$ denotes the post-collision RMs except the equilibrium. In
the above equations, $\bu_1 = \bu^2 -(\theta -1)\bdelta$ and $\bu_2 = \bu^2
-3(\theta -1)\bdelta$. The fourth-order post-collision term $\bbd{4,k}_p$ is
trimmed as the regularization applied in the previous work \cite{Li2019}. The
transformations are not limited to the first four orders, but it is sufficient
for the NSF hierarchy \cite{Li2019}. The post-collision distribution function
can be constructed by the Hermite expansion as
\begin{eqnarray}
	\bar{f}_{i,p} &=& \omega_i\sum_{n=0}^4\frac{1}{n!} \bm{a}_0^{(n)}: \bH{n}(\bxi_i) + \\ \nonumber
	&& \omega_i\sum_{n=1}^4\frac{1}{n!}\sum_k\bba{n,k}_p:\bS{n,k}(\bxi_i),
\end{eqnarray}

Then the stream can be conducted as
\begin{equation}
f_i(\bx+\bxi_i,t+1)= \bar{f}_{i,p}(\bx,t).
\end{equation}

\section{Numerical Simulation}

In this section, two numerical benchmarks are tested to verify the effectiveness
and accuracy of the present force scheme in the SMRT collision model. The first
one is the steady Talor-Green flow without boundary condition treatment and the
other one is the Womersley flow with unsteady force field and no-slip boundary
condition which is a non-trival issue for the high-order lattie LB model.
\label{section:VI} \subsection{Steady Taylor-Green flow} 
\begin{figure}[htbp]
\centering
\includegraphics[width=0.45\textwidth]{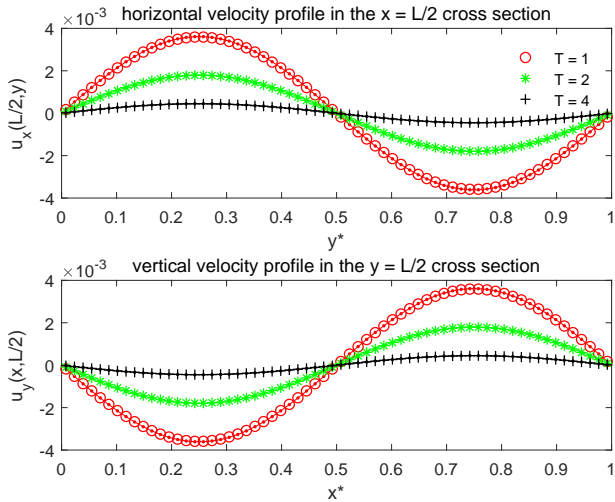}
\caption{Numerical (symbols,resolution:$64\times 64$) and analytical (solid
	line) results of 2D Taylor-Green flow at different times $t = nT$ with $n = 1,
	2, 4$ and $T=ln2/(2\nu k_1^2)$. }
\label{figure:TG_time}
\end{figure}

\begin{figure}[!htbp]
	\centering
	\includegraphics[width=0.5\textwidth]{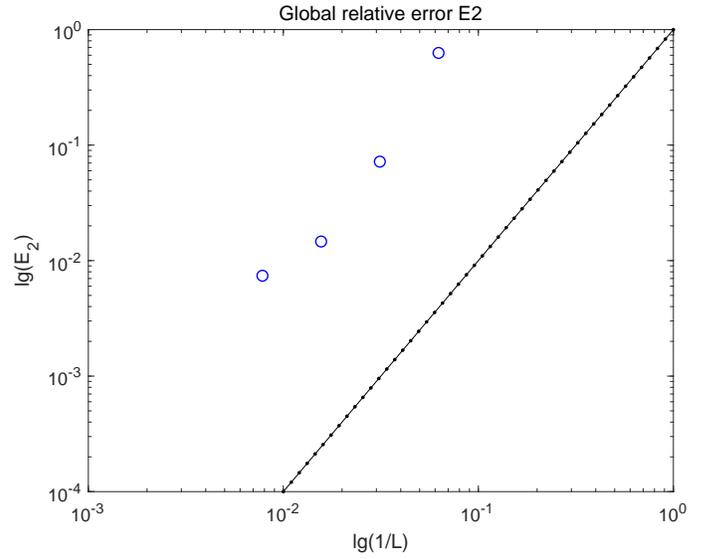}
	\caption{The convergence test: four resolutions are tested($16\times
	16,32\times 32,64\times 64,128\times 128$), the solid dot line is of slope 2,
	the convergence order of the present scheme is almost second-order.}
	\label{figure:convergence}
\end{figure}

For the two-dimensional
steady Taylor-Green flow within the periodic domain $[L,L]$, the analytical
unsteady force is exerted on the flow field
\begin{equation}
\bm{F}(x,y) = - \frac{\rho u_0^2}{2}\left[k_1 \sin\left( 2 k_1 x \right), \frac{k_1^2}{k_2} \sin\left( 2k_2y \right)\right]
\end{equation}
in which $k_1 = 2\pi /L,k_2 = 2\pi/L$,and $\nu$ is the kinematic viscosity, $u_0
= 0.005$ is the reference velocity. The flow field has the analytical solution
\begin{equation}
\bm{u}_a = - u_0 \left[\cos(k_1 x)\sin(k_2 y), - \frac{k_1}{k_2} \sin(k_1 x)\cos(k_2 y)\right]e^{t^*}, 
\end{equation}
in which $t^* = -2(k_1^2+k_2^2)\nu t$ and $t$ is the lattice unit. The flow is
characterized by the Reynolds number, $Re= u_0 L/\nu$. In the simulation, the
computational domain is resolved by a series of grid nodes,$L = [16,32,64,128]$,
with $Re = 50$. In Fig.~\ref{figure:TG_time}, the horizontal velocity profile
along the vertical center line and the vertical velocity profile along the
horizontal center line are depicted. It can be found the numerical simulation
results agree well with the analytical solution at different specific times.
Fig.~\ref{figure:convergence} depicts the global error with different
resolutions in contrast with the analytical solutions. The global error is
defined in Eq.(~\ref{equation:E2_error}). It can be found that the convergence
order is second order.
\begin{equation}
\label{equation:E2_error}
 E_2 = \sqrt{\frac{\sum \left( \bm{u} - \bm{u}_a \right)^2}{\sum  \bm{u}_a^2 }}
\end{equation}

\subsection{Womersley flow}

The second numerical benchmark is the Womersley flow. In this numerical case,
the flow is bounded by two parallel plate and a periodic pressure gradient or a
periodic force is exerted on the flow, which results in an unsteady flow. The
periodic force is $Fx(t) = -A_p \cos(\omega t)$, where $A_p$ is the amplitude
and $\omega = 2\pi/T$ is the frequency. Obviously, $Fx(t)$ is spatially uniform
but temporally unsteady. The analytical solution of the velocity field for the
Womersley flow is given by
\begin{equation}
U(y,t) = Re \left( i\frac{A_p}{\rho_0 \omega} \left\lbrace 1 - \frac{\cosh\left[ (1+i)y\sqrt{\frac{\omega}{2\nu}} \right] }{\cosh\left[ (1+i)L\sqrt{\frac{\omega}{2\nu}} \right]}  \right\rbrace  e^{ i\omega t} \right) 
\end{equation}
where  $y\in \left[ -L, L\right] $ with $2L$ being the channel width, $\nu$ is
the kinematic viscosity, and $Re$ denotes the real part of the complex number.

\begin{figure}[htbp]
\centering
\includegraphics[width=0.45\textwidth]{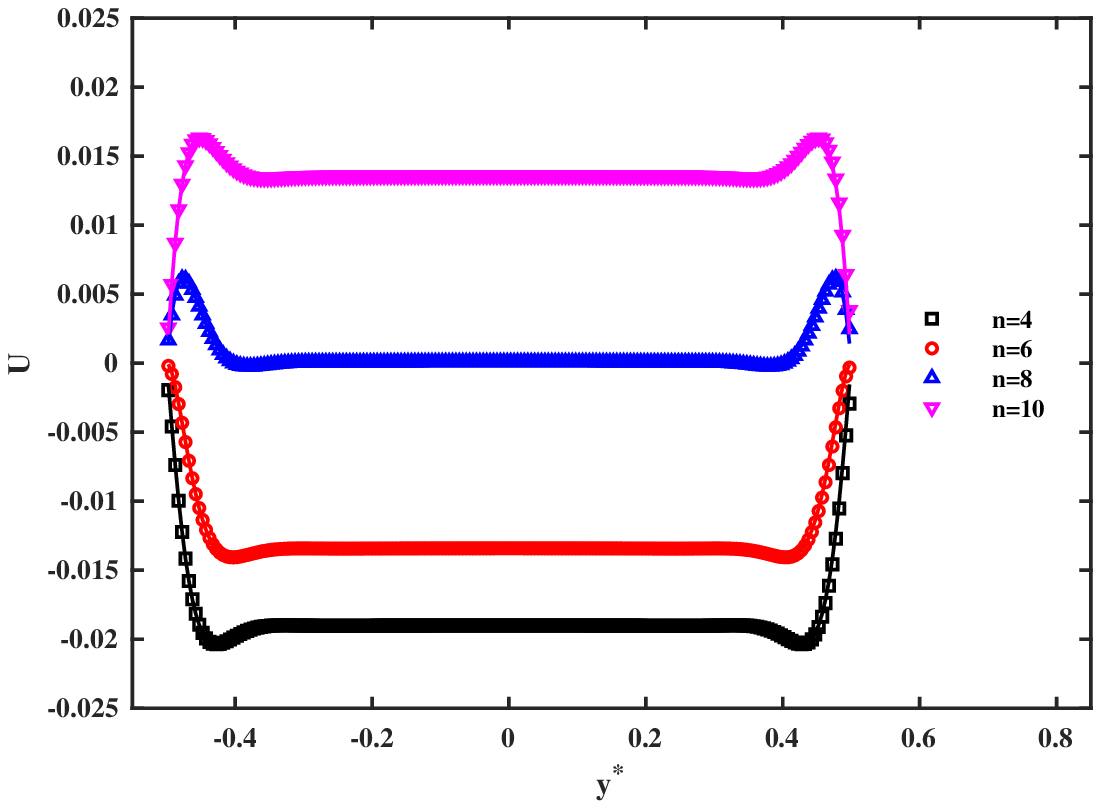}
\includegraphics[width=0.45\textwidth]{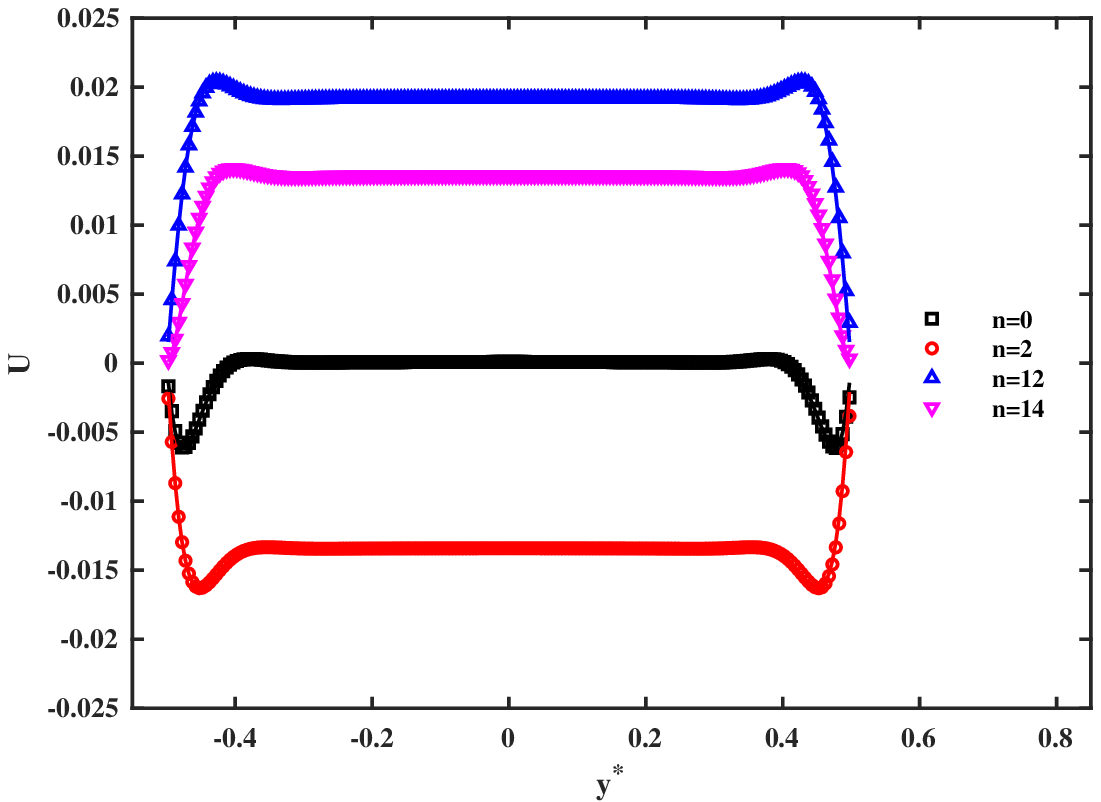}
\caption{Numerical (symbols) and analytical (solid line) results of
Womersley flow at different times $t = nT /16$ with (a) $n = 4, 6, 8, 10$
and (b) $n = 0, 2, 12, 14$. }
\label{figure:Womersley_comparison}
\end{figure}

The simulations are carried out in a computational domain with $Nx \times Ny =
50 \times 200$. In the $x$ direction, the periodic boundary condition is
applied. The diffuse reflection boundary condition~\citep{Meng2014} is imposed
on the two plates. The period $T$ is set as 1200, the kinematic viscosity is
chosen as $\nu = 0.1$, and the amplitude $Ap$ is set as 0.0001. The initial
density is chosen as $\rho = \rho_0 = 1$. The initial flow field is static. The
numerical results are obtained after running 20 periods. In
Fig.~\ref{figure:Womersley_comparison}, the velocity profile across two plates
at specific times are drawn in contrast with the analytical solutions. It can be
found that numerical results agree with the analytical solutions very well with
the unsteady force field.

\section{Conclusions and discussions}

\label{section:VII}

In this work, a generic \emph{a priori} derivation of the force scheme for
lattice Boltzmann method based on kinetic theoretical formulation is proposed
for the SMRT collision model. A second-order force scheme for the SMRT collision
model is obtained and numerically verified at isothermal level. This generic
approach of incorporating the force scheme actually can be applied to a wide
range of collision models. At the isothermal level, we found that the force term
is only not null at the first-order in the RCM space if the high-order
non-equilibrium terms are ignored. The new force scheme can account for body
force effects on high-order hydrodynamic moments which may be significant in
flows with high density and temperature gradients, which will be reported in the
following work.

\begin{acknowledgments}
This work was supported by the National Natural Science Foundation of China
Grants 51979053, 92152107, Natural Science Foundation of Heilongjiang Province
Grant LH2021A007, Heilongjiang Touyan Innovation Team Program, Department of
Science and Technology of Guangdong Province Grants 2019B21203001 and
2020B1212030001, and Shenzhen Science and Technology Program Grant
KQTD20180411143441009.
\end{acknowledgments}

\begin{appendices}
\section{Appendix: MRT force scheme in Raw-Moment Space}

In the raw-moment MRT model,the Galilean invariance and rotational invariance are not considered. For the convenient reading of this manuscript,
the derivation of the force scheme of the raw-moment MRT model is stated here. 
Similarly, starting from Eq.~(\ref{eq:dv}), we can get the same evolution equation as Eq.(~\ref{eq:evolution}) for the MRT model in the raw-moment space.
The transform of the right-hand side of Eq.(~\ref{eq:evolution}) to the RM space is
\begin{equation}
\label{eq:RHS_a}
\bar{\bm{a}}_p^{(n)} = \bar{\bm{a}}^{(n)} - \frac{1}{\tau_{n}}(\bm{a}^{(n)}-\bm{a}^{(n)}_0) + \bm{a}^{(n)}_F
\end{equation} 
in which the subscript \textit{p} denotes the post-collision state and $\bar{\bm{a}}^{(n)}$ is the mapping of Eq.~(\ref{eq:newf}) in RM space
\begin{equation}
\bar{\bm{a}}^{(n)} = \bm{a}^{(n)} + \frac{1}{2\tau_{n}}(\bm{a}^{(n)}-\bm{a}^{(n)}_0) -\frac{1}{2} \bm{a}^{(n)}_F
\end{equation}

Using the above equation, Eq.(~\ref{eq:RHS_a}) can be written as the following explicit and simple moment collision equation
\begin{equation}
\bar{\bm{a}}_p^{(n)} = \bm{a}_0^{(n)} +(1 - \frac{1}{\hat{\tau_{n}}})\bar{\bm{a}}_1^{(n)}  + (1- \frac{1}{2\hat{\tau_{n}}})\bm{a}^{(n)}_F,n\geq 1
\end{equation}
in which $\hat{\tau_{n}} = \tau_{n} + \frac{1}{2}$, and $\bar{\bm{a}}_1^{(n)} = \bar{\bm{a}}^{(n)} - \bm{a}_0^{(n)}$.

In the numerical implementation, $\bm{a}^{(n)}_F = n\bm{g}\bar{\bm{a}}^{(n-1)}$. If the non-equilibrium contributions are ignored, we have
\begin{subequations}
\label{eq:bbd1}
\begin{eqnarray}
\bm{a}^{(1)}_F &=& \bm{F}, \\ 
\bm{a}^{(2)}_F &=& 2\bm{F} \bm{u}  \\
\bm{a}^{(3)}_F &=& 3\bm{F}\left[\bm{u} \bm{u} + (\theta - 1) \bm{\delta} \right]  \\
\bm{a}^{(4)}_F &=& 4\bm{F}\left[\bm{u} \bm{u} \bm{u} + 3(\theta - 1)\bm{u} \bm{\delta} \right]  
\end{eqnarray}
\end{subequations}

In the RM collision operator, $\bar{\bm{a}}_1^{(1)} = -0.5\bm{F}$ which is
indicated by Eq.(~\ref{eq:half_F}). $\bar{\bm{a}}_1^{(n)}(n=2,3)$ is obtained by
the projection method. The fourth order term is obtained by the recursive
approach as $\bar{\bm{a}}_1^{(4)} = 4\bm{u}\bar{\bm{a}}_1^{(3)} -6\bm{u}_1
\bar{\bm{a}}_1^{(2)} + 4(\bm{u}\bm{u}_2) \bar{\bm{a}}_1^{(1)}$. Thus the
post-collision particle distribution function can be evaluated as
\begin{eqnarray}
	\bar{f}_{i,p} &=& \omega_i\sum_{n=0}^4\frac{1}{n!} \bm{a}_0^{(n)}: \bH{n}(\bxi_i) + \\ \nonumber
	& &\omega_i\sum_{n=1}^4\frac{1}{n!}\left\lbrace  (1 - \frac{1}{\hat{\tau_{n}}})\bar{\bm{a}}_1^{(n)}  + (1- \frac{1}{2\hat{\tau_{n}}})\bm{a}^{(n)}_F \right\rbrace : \bH{n}(\bxi_i) 
\end{eqnarray}

Then the stream can be conducted as
\begin{equation}
f_i(\bx+\bxi_i,t+1)= \bar{f}_{i,p}(\bx,t).
\end{equation} 

\end{appendices}

%


\end{document}